\documentclass{emulateapj} 
\pdfoutput=1
\usepackage{psfig} \usepackage{apjfonts}

\usepackage{graphicx}

\newcommand\beq{\begin{equation}}
\newcommand\eeq{\end{equation}}
\newcommand\lsim{\mathrel{\rlap{\lower4pt\hbox{\hskip1pt$\sim$}}
        \raise1pt\hbox{$<$}}}
\newcommand\gsim{\mathrel{\rlap{\lower4pt\hbox{\hskip1pt$\sim$}}
        \raise1pt\hbox{$>$}}}

\begin{document}
\title{Magnetic Drag on Hot Jupiter Atmospheric Winds}

\author{Rosalba Perna,\altaffilmark{1} Kristen Menou\altaffilmark{2,3}
and Emily Rauscher\altaffilmark{2,3}} \altaffiltext{1}{JILA and
Department of Astrophysical and Planetary Sciences, University of
Colorado, Boulder, CO, 80309} \altaffiltext{2}{Department of
Astronomy, Columbia University, 550 West 120th Street, New York, NY
10027} \altaffiltext{3}{Kavli Institute for Theoretical Physics, UCSB,
Santa Barbara, CA 93106}

\begin{abstract}

Hot Jupiters, with atmospheric temperatures $T \gsim 1000$~K, have
residual thermal ionization levels sufficient for the interaction of
the ions with the planetary magnetic field to result in a sizable
magnetic drag on the (neutral) atmospheric winds. We evaluate the
magnitude of magnetic drag in a representative three-dimensional
atmospheric model of the hot Jupiter HD 209458b and find that it is a
plausible mechanism to limit wind speeds in this class of atmospheres.
Magnetic drag has a strong geometrical dependence, both meridionally
and from the day to the night side (in the upper atmosphere), which
could have interesting consequences for the atmospheric flow
pattern. By extension, close-in eccentric planets with transiently
heated atmospheres will experience time-variable levels of magnetic
drag.  A robust treatment of magnetic drag in circulation models for
hot atmospheres may require iterated solutions to the magnetic
induction and Saha equations as the hydrodynamical flow is evolved.

\end{abstract}

\section{Introduction}

Hot Jupiters are close-in, presumably tidally-locked gaseous giant
planets orbiting only a few stellar radii away from their Sun-like
host star. By virtue of their slow rotation (synchronous with their
orbital periods, $\sim$ a few days), high atmospheric temperatures ($T
\gsim 1000$~K) and permanent day-side hemispheric forcing, hot
Jupiters are laboratories for the study of atmospheric dynamics in a
regime that is absent from the Solar System (see Showman et al. 2008
and Showman et al. 2010 for reviews). In recent years, considerable
progress has been made in observationally characterizing the
atmospheres of hot Jupiters via a combination of secondary eclipse,
transmission spectrum and phase curve measurements (see Deming 2008
and Charbonneau 2009 for reviews).  In parallel with this
observational progress, theoretical modeling of hot Jupiter
atmospheres has expanded greatly, in an attempt to provide robust
interpretations of the growing data set (see, e.g., Burrows \& Orton
2010, Showman et al. 2010 and Baraffe et al. 2010 for reviews).

One of the main interests in studying hot Jupiter atmospheres lies in
understanding their thermal and dynamical responses to the unusual
forcing conditions they are experiencing, with an atmospheric
circulation pattern that is likely different from anything known in
the Solar System. However, with this new regime also comes the
possibility that new physics is at play in these extreme atmospheres
(e.g., Menou \& Rauscher 2010). In this work, we investigate the
possibility that magnetic drag on atmospheric motions provides an
effective frictional mechanism limiting the asymptotic speeds of winds
in hot Jupiter atmospheres. This is particularly important as the fast
(transonic) speeds reached by winds in a variety of drag-free
atmospheric models for this class of planets (Dobbs-Dixon \& Lin 2008,
Dobbs-Dixon et al. 2010; Showman et al. 2009; Rauscher \& Menou 2010)
raise issues about compressibility, shocks and associated energy
conservation for the models (Goodman 2009; Rauscher \& Menou 2010).

Atmospheric motions are driven by pressure-gradient forces arising
from differential heating of the atmosphere. A small fraction of the
atmospheric "available" enthalpy is continuously converted into
kinetic energy of the atmospheric motions, which is itself
continuously dissipated by friction\footnote{In this context, friction
  typically refers to a dissipative process that is much more
  efficient than the microscopic viscosity of the atmospheric gas.}
(Lorenz 1955, Pearce 1978, Marquet 1991, Goodman 2009). In
steady-state, asymptotic wind speeds are thus reached through a
detailed balance between continuous thermal forcing and sustained
friction. While the source of wind friction on the Earth, and other
Solar System terrestrial planets by extension, is understood to be
largely associated with surface drag, the origin of friction in the
atmospheres of gaseous giant planets remains a major open question in
atmospheric science, even in the Solar System (e.g., Schneider \& Liu
2009; Liu et al. 2008; Showman et al. 2010). Identifying dominant
sources of internal friction in gaseous giant planet atmospheres can
thus be as important as adequately modeling their sources of thermal
forcing.

We show here that magnetic drag on weakly-ionized winds in the
predominantly neutral atmospheres of hot Jupiters, which arises from
wind interaction with the magnetic field generated in the planet's
bulk interior, is a plausible source of sizable friction, that may
need to be accounted for in atmospheric circulation models of hot
Jupiters. We note that, while this manuscript was being prepared,
Batygin \& Stevenson (2010) completed a study of the closely related
ohmic dissipation process associated with the currents induced by
magnetic drag. Here, we focus on the role of magnetic drag on
atmospheric winds and defer a study of ohmic dissipation to future
work.

\section{Atmospheric Currents and Ion drag}

\subsection{Model Atmosphere}

Our study is based on the three dimensional hot Jupiter atmospheric
circulation model of HD 209458b presented by Rauscher \& Menou
(2010). The model provides, at each point in the 3D atmosphere, values
for the pressure, temperature, zonal (east-west) and meridional
(north-south) wind velocity. Throughout this analysis, we use the
model atmosphere obtained after 1450 planetary days of
integration. The corresponding atmospheric structure and wind pattern
are described in detail in Rauscher \& Menou (2010).

The model extends vertically from 1 mbar at the top to 220 bar at the
bottom. Temperatures are typically around 1800 K in much of the deep
atmosphere, while they vary noticeably between day and night near the
model top, with lows of about 500~K and highs of about 1500~K.  From
the pressure and temperature, we infer local gas densities using the
ideal gas law, $\rho=\mu m_{\rm H}\,p/kT$, with mean molecular weight
$\mu=2.33 m_p$.  Densities in the atmosphere range from about
$10^{-3}$~g~cm$^{-3}$ at the model bottom to $\sim
10^{-8}-10^{-7}$~g~cm$^{-3}$ at the top.

\subsection{Ionization}

High in the atmosphere (say, at nanobar levels), UV photo-ionization is
important in determining the atmospheric ionization level (e.g.,
Murray-Clay et al. 2009), but in the relatively dense levels modeled
here, thermal ionization is expected to dominate the ionization
balance.\footnote{X-rays could provide a source of additional
  non-thermal electrons at and above mbar pressure levels, which is
  ignored from our calculation (J. Goodman, priv. communication).}

At the temperatures of interest in our model atmosphere, the main
source of free electrons is provided by thermal ionization of alkali
metals with low first-ionization potentials: Na, Al, and K.  Under
these conditions, the mean ion mass is on the order of $m_i\approx 30
m_p$ (see e.g. Draine et al. 1983).  For simplicity, we choose to
approximate Saha's equation for ionization balance with a formulation
that only accounts for the ionization of potassium (Balbus \& Hawley
2000),
\begin{eqnarray}
x_e\equiv\frac{n_e}{n_n}&=&6.47\times 10^{-13}\left(\frac{a_K}{10^{-7}}\right)^{1/2}
\left(\frac{T}{10^3}\right)^{3/4}\nonumber \\ 
&\times&\left(\frac{2.4\times 10^{15}}{n_n} 
\right)^{1/2}\frac{\exp(-25188/T)}{1.15\times 10^{-11}}\;,
\label{eq:xe}
\end{eqnarray}  
where $n_e$ and $n_n$ are the number densities of electrons and of
neutrals, respectively (in cm$^{-3}$), $a_K$ is the potassium
abundance, and $T$ is the temperature in K. Equation~(\ref{eq:xe}) is
a valid approximation only as long the resulting ionization fraction,
$x_e$, remains much smaller than the abundance of potassium, $a_K$. We
have verified that this condition is reasonably well satisfied for the
atmospheric conditions of interest here, with $x_e$ reaching at most
$\sim 10^{-9}$ in a few localized regions and taking much smaller
values ($x_e \sim 10^{-10}$-$10^{-14}$) in the rest of the atmosphere.
Throughout our analysis, we assume a near solar abundance of
potassium, $a_K = 10^{-7}$. This is not a critical model assumption
given the much more important, exponential dependence of $x_e$ with
temperature.  We also estimate that, even at $T \simeq 1800$~K, the
free electron contribution from sodium, which is $\sim 17$ times more
abundant than potassium at solar composition, approaches only
marginally that of potassium. While a more complete Saha equation
solution would be a clear improvement upon the very simple approach
adopted here, it would not qualitatively alter our main conclusions
about the role of magnetic drag.

\begin{figure*}
\centering \includegraphics[scale=0.45]{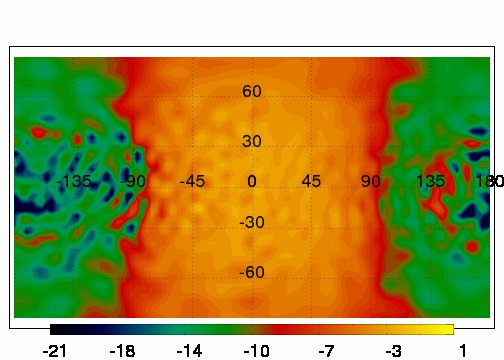}
\includegraphics[scale=0.45]{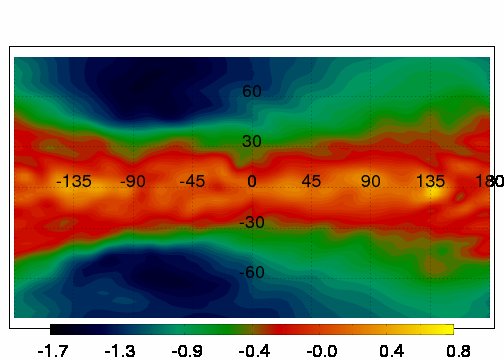}
\caption{Cylindrical maps of the magnetic Reynolds number, Log$_{10}
  (R_m)$, on two different pressure levels in the model
  atmosphere. The sub-stellar point is centered at longitude and
  latitude zero. {\em Left}: highest model level ($p=1$~mbar); {\em
    Right}: moderately deep level ($p=2$~bar). While very large
  day-night variations in $R_m$ occur high in the atmosphere, $R_m$
  values are much more uniform deeper in the atmosphere. At all
  levels, $R_m$ only marginally exceeds unity. }
\label{fig:Rm}
\end{figure*}

We assume conditions of gas neutrality, i.e. $n_e=n_i$, where $n_i$ is
the ionic number density. The coupling between charged $(i,e)$ and
neutral particles depends on their rate of collisions, $\left<\sigma
{\rm v}\right>_{i,e}$, for which we adopt the expressions from Draine et
al. (1983)
\begin{eqnarray}
\left<\sigma {\rm v}\right>_{i}&=&1.9\times 10^{-9}\;\; {\rm cm}^3 {\rm s}^{-1},\nonumber\\
\left<\sigma {\rm v}\right>_{e}&=&10^{-15}\left(\frac{128kT}{9\pi m_e}\right)^{1/2}{\rm cm}^3 {\rm s}^{-1}\;.
\end{eqnarray}
The electrical conductivity, $\sigma_e$, and the corresponding
resistivity, $\eta$, are respectively given by 

\beq \sigma_e =\frac{n_e e^2}{m_e n_n \left<\sigma {\rm v}\right>_e}\;\;\;{\rm
  and}\;\;\; \eta = \frac{c^2}{4\pi\sigma_e}\;.
\label{eq:sig-eta}
\eeq

\subsection{Non-ideal MHD regime}

Before proceeding with a calculation of induced currents, we need to
evaluate the relative importance of various non-ideal MHD effects.  We
begin by considering the full induction equation

\beq \frac{\partial {\bf B}}{\partial
  t}=\nabla\times\left[{\bf v}\times {\bf B } -\frac{4\pi \eta {\bf
      J}}{c}-\frac{{\bf J}\times {\bf B }}{c n_e} + \frac{\left({\bf
      J} \times {\bf B}\times \right) {\bf B}}{c\gamma \rho_i \rho}
  \right]\;,
\label{eq:ind}
\eeq where the last three terms are the ohmic, Hall and ambipolar
diffusion terms, respectively. In the above equation, the current
density \beq {\bf J}=\frac{c}{4\pi}\left(\nabla \times {\bf B}
\right)\;, \eeq ${\bf v}$ is the velocity of the neutrals, $\rho_i$ is
the ion mass density and $\gamma$ is the drag coefficient, which
is given by \beq \gamma = \frac{\left<\sigma {\rm v}\right>_{i}}{m_i + \mu}\;.
\eeq

We estimate the relative importance of the three non-ideal MHD terms
by selecting relevant velocity and length scales for the atmospheric
flow. For a representative induction velocity, we take the sound speed
$c_s$, since it is representative of the wind speeds achieved in the
modeled transonic flow.  Following Rauscher \& Menou (2009), we use the
adiabatic value $c_s=\sqrt(\Gamma kT/\mu m_p)$, with an adiabatic
index $\Gamma=1.47$. For now, we adopt a pressure scale height, $H_p$,
as a representative length scale for our comparison of the non-ideal
MHD terms, even though we argue below that using the resistive scale
height may be a more appropriate choice. With these choices, the
magnetic Reynolds number can be evaluated as

\beq R_m =\frac{c_s
  H_p}{\eta}\,,\;\;\;\;\;\;\;\; {\rm with}\;\;\; H_p=
\frac{R' \,T}{g}\,,
\label{eq:Rm}
\eeq 

where $R'=R/\mu=3.78\times 10^7$ erg/g/K is the effective gas constant
and $g=9.42$ m/s$^2$ is the gravitational acceleration of the planet.

We evaluate the relative magnitude of the Ohmic (Ohm), Hall (Hall) and Ambipolar
diffusion (Amb) terms with respect to the Induction term (Ind), using the
same notation as Balbus \& Terquem (2001; see also
Sano \& Stone 2002) for convenience, 
\begin{eqnarray} 
\frac{\rm Ohm}{\rm Ind} &\sim& \frac{\eta}{ {\rm v}
  L}=R_m^{-1}\,,\nonumber \\ 
\frac{\rm Hall}{\rm Ind} &\sim& \frac{cB}{4\pi e {\rm v} n_e L}\;,\nonumber \\
\frac{\rm Amb}{\rm Ind} &\sim&  \frac{B^2}{4\pi \gamma \rho\rho_i {\rm v}
  L}\,,
\label{eq:ratios}
\end{eqnarray}

with a velocity scale, v $=c_s$, and length scale, $L=H_p$.

We computed these ratios throughout the modeled atmosphere, assuming a
nominal field strength $B_0=3$~G, and found that the Ohmic term is
typically larger than the Hall term, by about 12 orders of magnitude
in the deepest levels, down to about 3 orders of magnitude in the
highest level modeled.  The Ohmic term is also found to dominate over
the Ambipolar diffusion term, although to a lesser extent. It is about
5 orders of magnitude larger in the deepest levels, down to a factor
of only a few in the uppermost modeled level.  We conclude that the
Hall term can be safely neglected throughout the modeled atmosphere,
while the Ambipolar diffusion term could in principle become important
in the more tenuous high atmosphere, at and above mbar pressure
levels, especially if magnetic field strengths $B \ga 10$~G are
considered. Based on these results and in the interest of simplicity,
we focus the rest of our analysis on the purely resistive MHD regime.

Figure 1 shows cylindrical maps of the magnetic Reynolds number on two
different pressure levels in our model atmosphere: the highest level,
at $p =1$~mbar (left panel), and a deeper level with $p \sim 2$~bar
(right). These magnetic Reynolds number maps largely trace the
variations in conductivity at a given pressure level. High in the
atmosphere, despite strong winds, the large temperature difference
between the day and the night side leads to many orders of magnitude
variations in $R_m$ values. By contrast, deeper in, temperature
variations on a given pressure level are modest and so are the
variations in $R_m$.

For the treatment of magnetic drag presented in \S\ref{ssec:cur} to be
valid, it is important that $R_m \ll 1 $ in the model
atmosphere. Indeed, an atmospheric flow with $R_m >1 $ is coupled to
the magnetic field well enough that, in principle, it could generate
its own magnetic field via dynamo action, leading to a situation that
could significantly complicate any discussion of magnetic
drag. Examining values of $R_m$ throughout our model atmosphere, we
find that they can exceed unity  in particular regions of the
atmosphere (as shown, e.g., in Figure~1), but only marginally so.

However, it is also likely that the simple definition of $R_m$ adopted
in Eq.~(\ref{eq:Rm}) effectively overestimates the ability of the
model atmosphere to generate its own dynamo.  Indeed, the length scale
for dimensional analysis in a geometrically anisotropic system like an
atmosphere must be chosen carefully.  Examination of the specific
toroidal component of the induction equation used below (see
Eq.~[\ref{eq:indphi}]) shows that vertical gradients generally
dominate over latitudinal ones. The choice of the vertical gradient
scale itself is subtle since it is different for pressure ($H_p$) and
resistivity ($H_\eta$), with a generally shorter resistivity scale
from its exponential dependence on temperature. Based on the leading
order vertical gradient terms entering Eq.~(\ref{eq:indphi}), this
suggests that $H_\eta$ may be the relevant scale to define $R_m$ in
our problem, rather than $H_p$. Indeed, $H_\eta$ is generally smaller
than $H_p$ in our model atmosphere, by a factor of a few typically on
the dayside and a factor $\sim 10$ to 100 on the nightside in the
upper levels, and by a factor of a few deeper in.  Based on these
arguments, we assume for the remainder of this analysis that there is
no self-induced atmospheric dynamo, so that the planetary magnetic
field, originating from deep currents in the bulk interior, is the
only relevant one.

\subsection{Induced currents and magnetic drag} \label{ssec:cur}

Since our goal in this work is to establish whether or not the typical
magnitude of magnetic drag in hot Jupiter atmospheres is sufficient to
appreciably affect their circulation patterns, we adopt the simplest
approach possible and make a number of simplifying assumptions.  We
assume the planetary magnetic field to have a dipolar geometry, with a
dipole axis coincident with the rotation axis.  The surface magnetic
field strengths expected for hot Jupiters are rather
uncertain. Jupiter has a field strength varying from about 4 G at the
equator to about 15 G at the poles. In our calculations, we assume a
fiducial surface field strength $B_0=3$ G (at the magnetic pole) and
discuss the effects of larger field values whenever appropriate.

We adopt a formalism that is closely related to that described by Liu
et al. (2008). In particular, we strictly focus our analysis on the
zonal (= east-west) component of the atmospheric flow and the drag
experienced by this flow as it generates a toroidal field component,
and associated currents, from the purely poloidal planetary magnetic
field. The generation of the toroidal component of the magnetic field
is governed by the resistive induction equation
\begin{eqnarray}
\frac{\partial B_\phi}{\partial t} &=& r\sin\theta\left[\frac{\partial\Omega}{\partial r}  
B_r + \frac{1}{r}\frac{\partial\Omega}{\partial \theta}  
B_\theta\right] + \frac{1}{r}\frac{\partial}{\partial r}\left[\eta\frac{\partial}{\partial r}
(rB_\phi)\right] \nonumber \\
&+& \frac{1}{r^2}\frac{\partial}{\partial\theta}\left[\frac{\eta}{\sin\theta}
\frac{\partial}{\partial\theta}(\sin\theta B_\phi)\right]\;,
\label{eq:indphi}
\end{eqnarray}
where $\Omega={\rm v}_\phi r^{-1}\sin^{-1}\theta$ in spherical coordinates
($r$, $\theta$, $\phi$). We seek a steady-state solution to this
induction equation on the assumption that it represents a reasonable
value for the level of drag expected in our model atmosphere with
prescribed wind speeds. For simplicity, we entirely neglect the
poloidal component of the induction equation, which would lead to
magnetic drag on the meridional (=north-south) component of the
circulation. While this assumption may be justifiable in much of the
model atmosphere, where the dominant circulation is zonal, it is
likely a poor approximation high up in the atmosphere, where the
circulation pattern is away from the substellar point and thus equally
meridional as zonal.

\begin{figure}
\centering
\includegraphics[scale=0.45]{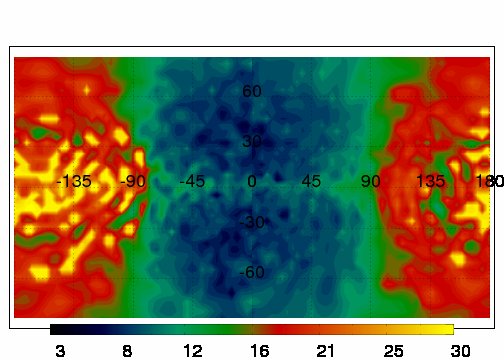}\\
\includegraphics[scale=0.45]{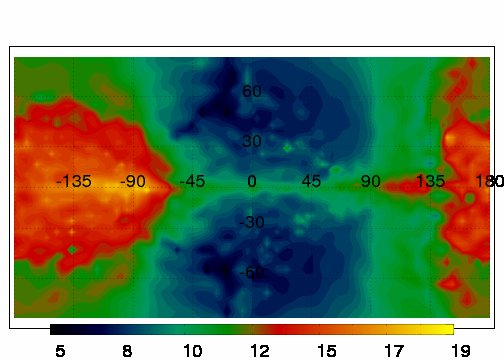}\\
\includegraphics[scale=0.45]{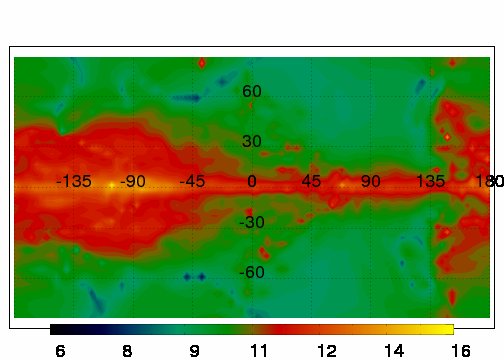}\\
\includegraphics[scale=0.45]{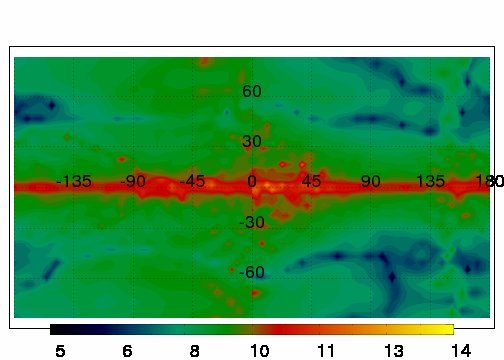}
\caption{Cylindrical maps of the zonal drag time, Log$_{10}[\tau_{\rm
      drag}~(s)]$, on four different pressure levels in the model
  atmosphere. The sub-stellar point is centered at longitude and
  latitude zero. {\em From top to bottom}: highest model level, at $p=1$~mbar;
  $p=40$~mbar level;  $p=0.5$~bar
  level; and last a moderately deep level, at $p=2$~bar.
  High in the atmosphere (top two panels), drag times vary very strongly
  between the day and the night sides.  Drag times also become very
  long in the vicinity of the equator, by virtue of the aligned
  dipolar field geometry adopted.}
\label{fig:tdrag}
\end{figure}

In our three-dimensional model atmosphere, resistivity varies strongly
in the vertical but it also does so horizontally, in the upper
atmosphere (see, e.g., the left panel of Fig.~1). It is thus not a
priori clear whether the second-to-last term dominates over the last
term in Eq.~(\ref{eq:indphi}). The relative magnitude of these two
terms depends on the relative magnitude of the vertical gradient scale
$H_\eta(r)=\left|\eta(r,\theta,\phi)
/(d\eta(r,\theta,\phi)/dr)\right|_{\theta,\phi}$ and the horizontal
gradient scale $H_\eta(\theta)=(1/r)\,\left|\eta(r,\theta,\phi)
/(d\eta(r,\theta,\phi)/d\theta)\right|_{r,\phi}$ for resistivity.  We
computed the ratio $H_\eta(r)/H_\eta(\theta)$ in our model atmosphere
and found it to be $<0.1$ over the vast majority of the atmospheric
domain. As a result, the last term in Eq.~(\ref{eq:indphi}) is small
and can be dropped from the analysis.  The domain in which horizontal
resistivity gradients can approach vertical gradient values is about
1\% of the total, and about 2/3 of this domain resides in the model
uppermost layers, at $p \la 10$~mbar.  For these upper levels, our
estimates for the magnetic drag may thus be inaccurate in some regions
because the strong horizontal resistivity gradients can no longer be
ignored.

Under the various assumptions made so far, the meridional component of
the current induced by the zonal flow is dominant and can be evaluated
as (Liu et al. 2008):
\begin{eqnarray}
j_\theta(r,\theta,\phi)&=&-\frac{c \sin\theta}{4\pi r\eta(r,\theta,\phi)}
\int_r^R dr' {r'}^2\left( \frac{\partial \Omega}{\partial r'}B_r  
+\frac{1}{r'}\frac{\partial \Omega}{\partial\theta} B_\theta\right)\\
&+& \frac{R\eta(r,\theta,\phi)}{r\eta(R,\theta,\phi)} j_\theta(R,\theta,\phi)\;,
\label{eq:j}
\end{eqnarray}
where the last term is associated with a boundary current in the
uppermost modeled level, $j_\theta(R,\theta,\phi)$. Eq.~(\ref{eq:j})
expresses that the local current scales as an integral of the zonal
flow above the level of interest. Lacking information about the
nature of currents possibly flowing from regions above the modeled
atmospheric layers, we set this boundary current to zero for
simplicity. This represents an important source of uncertainty in our
modeling approach, but we note that, unless near cancellations occur,
additional boundary currents could in principle contribute to even
stronger magnetic drag than estimated here.

We evaluate the magnitude of the magnetic drag on the (mostly neutral)
flow by using a standard formulation for ``ion drag'' based on the
bulk Lorentz force experienced by the ionic component (see, e.g., Zhu
et al. 2005),

\beq
\rho\frac{d {\bf v}}{d t} \propto \frac{1}{c}\;
         {\bf j}\times {\bf B}\;,
\label{eq:dvdt}
\eeq where ${\bf j}$ is the current induced in the atmosphere by the
zonal flow. From this, we deduce the typical zonal drag time over
which the zonal flow would be brought to a halt, \beq \tau_{\rm drag}
\sim \frac{\rho\, \left|{\rm v}_{\phi}\right| \,c}{|{\bf j_\theta}\times
  {\bf B }|}\;,
\label{eq:tdrag}
\eeq in the absence of any other forces.

An order of magnitude estimate for the drag can be simply derived by
making the approximation $j_\theta\sim c{\rm v}_\phi B/(4\pi\eta)$,
which yields
\beq
\tau_{\rm drag}
\sim \frac{4 \pi\rho\,\eta}{B^2\cos\theta}\sim 10^8\, 
\frac{\rho_{-5}\,\eta_{13}}{{B_3}^2 \cos\theta}\, {\rm s}\;,
\eeq
where $B_3\equiv B/(3\,{\rm G})$, $\rho_{-5}\equiv \rho/(10^{-5}\,{\rm g}\,{\rm cm}^{-3})$,
$\eta\equiv \eta/(10^{13}\,{\rm cm}^2 \,{\rm s}^{-1})$ are typical values, 
and $\pi - \theta$ is the angle between ${\bf j}_\theta$ and ${\bf B}$. 

Fig.~2 displays maps of the magnetic drag time from the detailed
modeling of Eq.~(\ref{eq:tdrag}), at four depths within the
atmosphere.  At deeper levels, the drag time is relatively long,
$\tau_{\rm drag}\ga 10^7-10^8$ sec, with only a few localized pockets
with shorter drag times, corresponding to regions of the flow with low
velocities.  Higher in the atmosphere, the magnetic drag time spans a
very large range of values, with very long drag times on the nightside
and considerably shorter drag times on the dayside.  While magnetic
drag is probably negligible on the night side, the short drag times
$\sim 10^4-10^6$ sec on the day side suggest that magnetic drag could
be dynamically important in these regions. In addition to the
day-night asymmetry in drag times high up in the atmosphere, Fig.~2
shows that zonal drag times are consistently long along the
equator. This geometrical effect, which results from the aligned
dipolar field geometry assumed in our analysis (leading to $B_r =0$ at
the equator), could have interesting consequences for the atmospheric
flow, such as promoting barotropic shear instabilities in the
equatorial regions (see, e.g., Menou \& Rauscher 2009). This
geometrical effect may persist for dipolar field geometries with a
small amount of misalignment.

\begin{figure*}
\centering 
\includegraphics[scale=0.38]{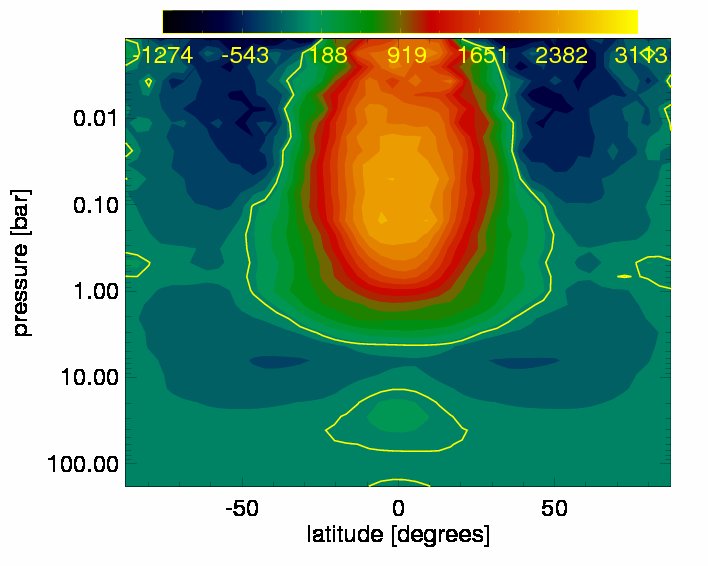}
\includegraphics[scale=0.38]{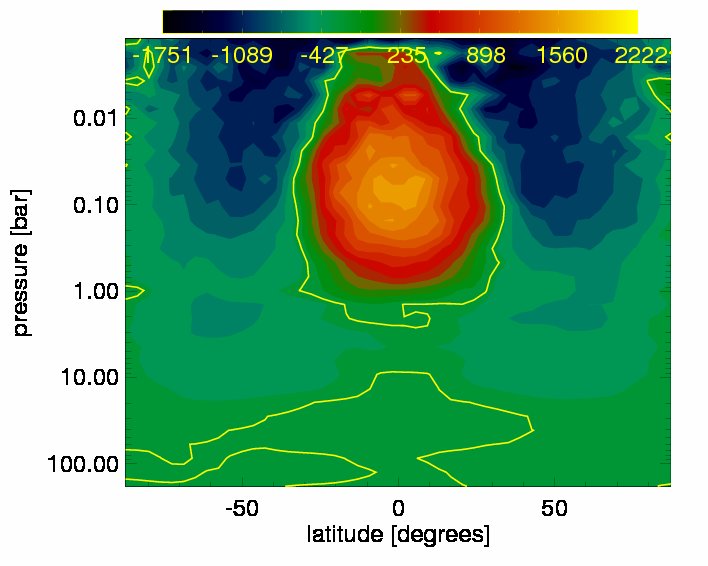}
\includegraphics[scale=0.38]{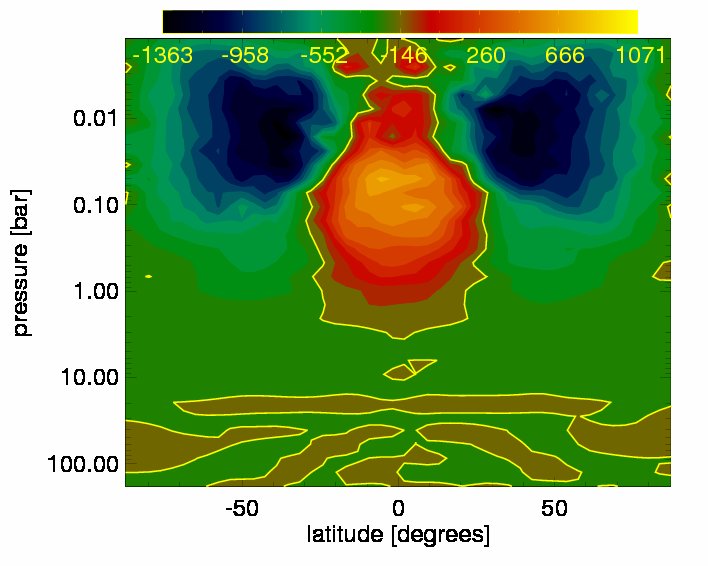}
\caption{Zonal average of the zonal wind in m~s$^{-1}$, as a function
  of latitude and depth, in two atmospheric circulation models with
  vertically-varying and horizontally-uniform levels of Rayleigh
  drag. Yellow lines separate regions of positive (eastward) flow and
  negative (westward) flow. These contour plots are meant to be
  compared to the equivalent version for the drag-free model described
  in Rauscher \& Menou (2010).  {\em Top}: Rayleigh drag times
  calibrated to match the typical magnetic drag times expected for a
  field strength $B \sim 3$~G; {\em Middle}: Rayleigh drag times
  calibrated to match a magnetic field strength $B \sim 10$~G. {\em
    Bottom}: Rayleigh drag times calibrated to match a magnetic field
  strength $B \sim 30$~G.  Significant modifications to the
  atmospheric flow emerge, relative to the drag-free case,
  particularly in terms of the zonal-mean wind speeds (see text for
  details).}
\label{fig:ZU}
\end{figure*}

\subsection{Preliminary models with Rayleigh drag}

Since the shortest drag times shown in Fig.~2 are comparable to a
representative zonal advection time in our model atmosphere,
$\tau_{\rm adv} \sim R_p/c_s \sim 5 \times 10^4$~s, one may expect
such levels of magnetic drag to affect the atmospheric flow
non-trivially, even in the presence of continuous thermal forcing of
the circulation.

To test this hypothesis, we have run modified versions of the
drag-free model described in Rauscher \& Menou (2010) in which a
vertically-varying but horizontally-uniform level of Rayleigh drag
($\propto - {\bf v}/\tau_{\rm drag}$; see Menou \& Rauscher 2009) is
applied throughout the modeled flow. For simplicity, an equal amount
of drag is applied to the zonal and the meridional velocity fields and
only average, representative values of $\tau_{\rm drag}$ for each
pressure level are employed. Assuming a field strength $B_0=3$~G, we
evaluated representative horizontally-averaged values of $\tau_{\rm
  drag} \simeq 6 \times 10^6$~s at 1 mbar (model top), $\simeq 4
\times 10^7$~s at 40~mbar, $\simeq 7 \times 10^7$~s at 2~bar and
$\simeq 8 \times 10^8$~s at 220~bar (model bottom), with a logarithmic
dependence of $\tau_{\rm drag}$ on pressure in between.

We have run models for the nominal field strength $B_0 =3$~G as well
as models with drag times uniformly reduced by a factor 10 and 100
(corresponding to field strengths $B \sim 3 \times B_0$ and $10 \times
B_0 $). Each model was run for 500 planet days (=rotation periods). In
the model with $B = B_0$, a very similar atmospheric flow structure to
that described in Rauscher \& Menou (2010) is obtained, with zonal
wind speeds slightly reduced at the $\sim 5$--$10\%$ level
(Fig.~\ref{fig:ZU}, top panel). In the model with $B = 3 \times B_0$,
the atmospheric flow is again qualitatively similar but the
superrotating equatorial jet is shallower and confined to levels above
$\sim 1$~bar, while zonal wind speeds are typically reduced by a
factor $\sim 30\%$ (Fig.~\ref{fig:ZU}, middle panel). Finally, in the
model with $B = 10 \times B_0$, we witness significant changes in the
structure of the atmospheric flow. Zonally-averaged zonal wind speeds
in the equatorial jet are reduced by a factor $\sim 3$ while stronger
counter jets dominate at mid-latitudes (Fig.~\ref{fig:ZU}, bottom
panel). Nevertheless, peak velocities in the upper atmosphere remain
supersonic.  While these three simple runs are only illustrative,
since they do not differentiate zonal and meridional drag or include
the strong horizontal dependence of the drag (Fig.~2), they clearly
suggest that magnetic drag of the typical magnitude evaluated in this
work can have a noticeable, and possibly important, impact on the
atmospheric flow of hot Jupiters.

\section{Summary and Discussion}

Circulation models for hot Jupiter atmospheres have gained substantial
interest in recent years thanks to the direct detections of such
atmospheres, which indicate a possible role for winds in shaping the
emergent properties of these planets.  All existing calculations have
been purely hydrodynamical in nature, neglecting the possibility that
interactions between the flow and the planetary magnetic field could
influence the circulation. However, given the high atmospheric
temperatures of these planets, the small fraction of the fluid that is
ionized may be sufficiently large to effectively couple the mostly
neutral flow to the planetary magnetic field.

In this work, we have evaluated the typical magnitude of the magnetic
drag exerted on hot Jupiter atmospheric winds, using the flow pattern
obtained in the 3D circulation model of Rauscher \& Menou (2010). Our
results indicate that typical magnetic drag stopping times are a
strong function of depth, longitude and latitude.  In the upper model
atmosphere, temperature differences between the day and night side
cause extreme variations of the magnitude of the drag, which could be
dynamically important only on the day side.  At all levels, zonal drag
also varies strongly with latitude in the vicinity of the equator, by
virtue of the aligned dipole field geometry adopted.

By adopting a simplified treatment of ionization balance, focusing on
the zonal component of drag only, ignoring upper boundary currents or
the role of ambipolar diffusion, and neglecting horizontal variations
in resistivity, the calculations presented in this work are clearly of
limited scope. Nevertheless, they appear sufficient to make the case
that dynamically interesting levels of magnetic drag may have to be
accounted for in circulation models for hot Jupiter atmospheres. This
conclusion is supported by a simple comparison between a drag-free
circulation model and a few additional models with Rayleigh drag
applied at levels commensurate with that expected from magnetic drag,
which show dynamically interesting consequences for the flow,
especially for large magnetic field values.

A better assessment of the role of magnetic drag in hot Jupiter
atmospheres may require iterated solutions to the multi-dimensional
induction equation, together with Saha's equation, as the
time-dependent hydrodynamical flow is evolved. Such work will be
important to re-evaluate the role of deep ohmic dissipation in
possibly inflating hot Jupiters (Batygin \& Stevenson 2010) and to
elucidate the plausible diversity of atmospheric behaviors expected
from a variety of field strengths and geometries. The strong
dependence of resistivity on atmospheric temperature could, in
principle, lead to different classes of atmospheric behaviors as a
function of mean orbital separation, and it will likely lead to an
interesting phenomenology for eccentric close-in planets, as they
experience time-variable levels of magnetic drag (and ohmic
dissipation).

\acknowledgements 

We thank Adam Burrows, Jeremy Goodman and Peter Goldreich for useful
discussions and encouragements.


\begin{references}


\reference{}
Balbus, S. A., Hawley, J. F. Space Science Rev., 92, 39

\reference{}
Balbus, S., A., Terquem, C. 2001, ApJ, 552, 235

\reference{} Baraffe, I., Chabrier, G., Barman, T. 2010, Reports on
Progress in Physics, Vol. 73, Issue 1, pp. 016901

\reference{} 
Batygin, K., Stevenson, D.  J. 2010, submitted to ApJL, eprint arXiv:1002.3650

\reference{} Burrows, A., Orton, G. to appear in
``Exoplanets'', Spring 2010
Space Science Series of the University of Arizona Press (Tucson, AZ);
Ed. S. Seager; eprint arXiv: 0910.0248

\reference{} Charbonneau, D., in ``Transiting Planets'', Proceedings of
the International Astronomical Union, IAU Symposium, Volume 253,
p. 1-8

\reference{}
Deming, D. 2008, to appear in Proceedings of IAU Symposium 253; eprint arXiv:0808.1289
\reference{}

\reference{}
Dobbs-Dixon, I.,  Lin, D. N. C. 2008, ApJ, 673, 513

\reference{}
Dobbs-Dixon, I., Cumming, A., Lin, D. N. C. 2010, ApJ, 710, 1395

\reference{}
Draine, B. T., Roberge, W. G., Dalgarno, A. 1983, ApJ, 264, 485

\reference{}
Goodman, J. 2009, ApJ, 693, 1645

\reference{}
Liu, J., Goldreich, P. M., Stevenson, D. J. 2008, Icar., 653, 664

\reference{}
Lorenz, E. N. 1955, Tell, 7, 157

\reference{}
Marquet, P. 1991,  Quart. J. Roy. Meteo. Soc. 117, 449

\reference{}
Menou, K., Rauscher, E., 2010, ApJ in press, eprint ArXiv:0910.1346


\reference{}
Menou, K., Rauscher, E., 2009, ApJ 700, 887


\reference{}
Murray-Clay, R. A., Chiang, E. I., Murray, N. 2009, ApJ, 693, 23

\reference{}
Pearce, R. P. 1978, Quart. J. Roy. Meteo. Soc. 104, 737

\reference{}
Rauscher, E. Menou, K. 2010, submitted to ApJ, eprint arXiv:0907.2692

\reference{}
Sano, T., Stone, J., M. 2002, ApJ, 577, 534

\reference{}
Schneider, T., Liu, J.  2009, JAtS, 66, 579

\reference{}
Showman, A. P., Y-K. Cho, J., Menou, K. 2010, to appear in
``Exoplanets'', Spring 2010
Space Science Series of the University of Arizona Press (Tucson, AZ);
Ed. S. Seager; eprint arXiv: 0911.3170.

\reference{}
Showman, A. P. et al. 2009, ApJ 699, 564

\reference{} 
Showman, A. P., Menou, K., Cho, J. Y.-K. 2008, in ``Extreme
Solar Systems'', ASP Conference Series, Vol. 398, proceedings of the
conference held 25-29 June, 2007, at Santorini Island, Greece. Edited
by D. Fischer, F. A. Rasio, S. E. Thorsett, and A. Wolszczan, p.419

\reference{} 
Zhu, X., Talaat, E. R., Baker, J. B. H. \& Yee, H.-H. 2005, Ann. Geo.
2005, 23, 3313

\end{references}
\end{document}